\documentclass{nature}
\usepackage{xpatch}

\newcommand\ulap[1]{\vbox\@to\z@{{\vss#1}}}%
\newcommand\dlap[1]{\vbox\@to\z@{{#1\vss}}}%

\usepackage{graphicx}
\usepackage{tabularx}
\usepackage{multirow}
\usepackage{gensymb}

\usepackage{amssymb}
\usepackage[version=4]{mhchem}

\newcommand\arcsec{\mbox{$^{\prime\prime}$}}%

\usepackage[dvipsnames]{xcolor}

\def\farcm@apj{%
 \mbox{.\kern -0.7ex\raisebox{.9ex}{\scriptsize$\prime$}}%
}%

\usepackage{lineno}

\title{Polarized thermal emission from dust in a galaxy at redshift 2.6}

\author{J.\,E.~Geach$^1$, E.~Lopez-Rodriguez$^2$, M.\,J.~Doherty$^1$, Jianhang~Chen$^3$, R.\,J.~Ivison$^{3,4,5,6}$, G.\,J.~Bendo$^7$, S.~Dye$^8$, K.\,E.\,K.~Coppin$^1$}

\begin{document}
\bibliographystyle{naturemag}

\maketitle

\begin{affiliations}
\small
 \item Centre for Astrophysics Research, School of Physics, Engineering and Computer Science, University of Hertfordshire, Hatfield, UK
 \item Kavli Institute for Particle Astrophysics \& Cosmology, Stanford University, Stanford, CA, USA
 \item European Southern Observatory, Garching, Germany
 \item Department of Physics and Astronomy, Macquarie University, Sydney, New South Wales, Australia
 \item School of Cosmic Physics, Dublin Institute for Advanced Studies, Dublin, Ireland
 \item Institute for Astronomy, Royal Observatory, University of Edinburgh, Edinburgh, UK
 \item UK ALMA Regional Centre Node, Jodrell Bank Centre for Astrophysics, Department of Physics and Astronomy, The University of Manchester, UK
 \item School of Physics and Astronomy, University of Nottingham, Nottingham, UK
\end{affiliations}

\begin{abstract}
Magnetic fields are fundamental to the evolution of galaxies, playing a key role in the astrophysics of the interstellar medium (ISM) and star formation. Large-scale ordered magnetic fields have been mapped in the Milky Way and nearby galaxies\cite{Beck2019, Lopez-Rodriguez_2023}, but it is not known how early in the Universe such structures form\cite{Mao17}. Here we report the detection of linearly polarized thermal emission from dust grains in a strongly lensed, intrinsically luminous galaxy that is forming stars at a rate more than a thousand times that of the Milky Way at redshift 2.6, within 2.5 Gyr of the Big Bang\cite{Geach2015,Geach2018}. The polarized emission arises from the alignment of dust grains with the local magnetic field\cite{Hoang2008,Andersson2017}. The median polarization fraction is of order one per cent, similar to nearby spiral galaxies\cite{Lopez-Rodriguez2022}. Our observations support the presence of a 5\,kiloparsec-scale ordered magnetic field with a strength of around 500$\mu$G or lower, orientated parallel to the molecular gas disk. This confirms that such structures can be rapidly formed in galaxies, early in cosmic history. 

\end{abstract}
   
We observed the lensed galaxy 9io9\cite{Geach2015} with the Atacama Large Millimeter/Submillimeter Array (ALMA) at a representative frequency of 242\,GHz (equivalent to a wavelength of approximately 350\,$\mu$m in the rest-frame of the galaxy) to record the dust continuum emission averaged over a total bandwidth of 7.5\,GHz. The set of {\it XX}, {\it YY}, {\it XY} and {\it YX} linear polarization parameters recorded in full polarization mode allow measurement of the Stokes parameters $Q$ and $U$, yielding the total linearly polarized intensity, $PI=\sqrt{Q^2 + U^2}$, and position angle (PA) of polarized emission $\chi = 0.5\arctan{(U/Q)}$. The root mean squared sensitivity of the observations is $\sigma_{I} = 47\,\mu\text{Jy}\,\text{beam}^{-1}$ and $\sigma_{Q} \sim \sigma_{U} = 9\,\mu\text{Jy}\,\text{beam}^{-1}$. In Figure~1 we present image plane maps of the total intensity $I$, Stokes $Q$ and $U$, and polarized intensity $PI$. The polarization angle $\chi$ is rotated by 90\,degrees to show the plane-of-the-sky magnetic ({\it B}) field orientation ($\chi_{B}$). We measure an image plane integrated flux density of $I=62$\,mJy, integrated polarization fraction of $P=0.6\pm 0.1$\%, where $P=PI/I$, and {\it B}-field orientation of $\chi_{B} = (-0.7\pm1.4)$\,degrees. The mean of the distribution of polarization fractions and {\it B}-field orientations is $\langle P \rangle = 0.6\pm0.3$\% and $\langle \chi_{B} \rangle = (0.8 \pm 18.3)$\,degrees, respectively. Note that the uncertainties are the dispersion of the distribution of individual measurements within the galaxy, not the accuracy in the polarization measurement (Methods).   

\begin{figure}
    \centering
    \includegraphics[width=\textwidth]{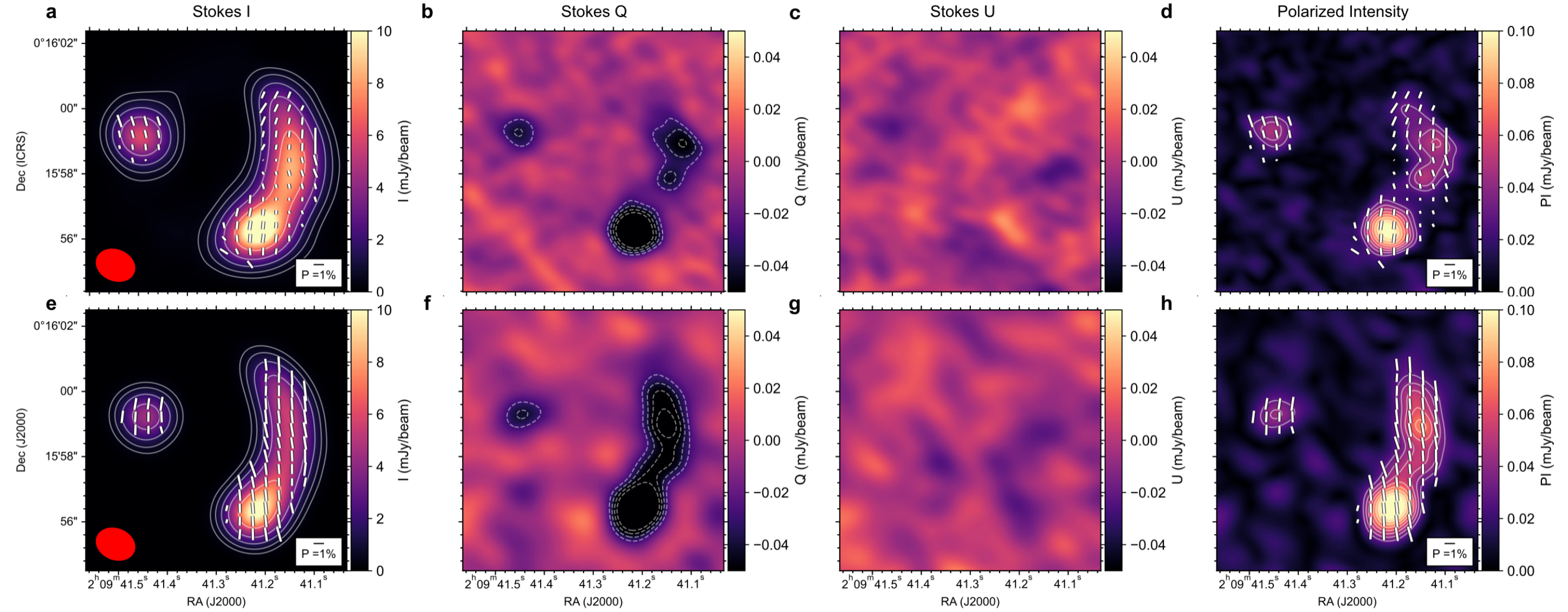}
    \caption{\textbf{The magnetic field orientation of the gravitationally lensed galaxy 9io9 at z=2.553.} 
    (\textbf{a}-\textbf{d}), ALMA $242$ GHz polarimetric observations of the Stokes $I$, $Q$, and $U$ parameters, and the polarized intensity $PI$. The synthetic beam of the observations ($1.2\arcsec\times0.9\arcsec$, $\theta=68$\,degrees) is shown as the red ellipse, lower left. The {\it B}-field orientation is indicated by white lines displayed at the Nyquist sampling, with line lengths proportional to the polarization fraction. 
    (\textbf{e}-\textbf{h}),  Synthetic polarimetric observations using a constant {\it B}-field configuration in the source plane. Contours indicate signal-to-noise: for Stokes $I$, the contours increase as $\sigma_{I} \times 2^{3,4,5,\dots}$. For Stokes $Q$ and $U$, and $PI$, the contours start at $3\sigma$ and increase in steps of $1\sigma$.}
    \label{fig:fig1}
\end{figure}

Using the lens model derived from previous high-resolution millimetre continuum emission and optical {\it Hubble Space Telescope} imaging, the source plane CO(4--3) emission, tracing the cold molecular gas reservoir, has been shown to be well-modelled by a rotating disk of maximum radius 2.6\,kpc, inclined by approximately 50\,degrees to the line of sight, with a position angle (PA) on the sky of approximately 5\,degrees East of North\cite{Geach2015,Geach2018}. With this model as a constraint, we explore what source plane {\it B}-field configurations are consistent with the image plane polarization observations. The most likely source plane configuration is a large-scale ordered {\it B}-field orientated $\chi_{\rm B} =5^{+5}_{-10}$\,degrees east of north with an extent matching that of the CO emission (Figure~2). This result implies the presence of a 5\,kpc-scale galactic ordered magnetic field orientated parallel to the molecular gas-rich disk. Angular variations of $\chi_{\rm B}$ across the galaxy present in the image plane maps, corresponding to scales of 600\,pc in the source plane, can be explained by the low signal-to-noise ratio and beam effects (Methods). This result implies that the introduction of a random {\it B}-field component with an angular variation of $\pm5$\,degrees in addition to the large-scale ordered \textit{B}-field is also consistent with the observations. We currently lack the sensitivity and resolution to map the  configuration of the {\it B}-field strength at scales $\sim$100\,pc where structure related to turbulence can start to be resolved. The observed \textit{B}-field configuration parallel to the disk is consistent with the galactic \textit{B} fields measured in local spiral galaxies observed at far-infrared and radio wavelengths\cite{Beck2019, Lopez-Rodriguez_2023}. Note that our far-infrared polarimetric observations trace  a density-weighted average \textit{B}-field in the cold and dense ISM, rather than a volume average \textit{B}-field in the warm and diffuse ISM by radio polarimetric observations.

\begin{figure}
    \centering
    \includegraphics[width=\textwidth]{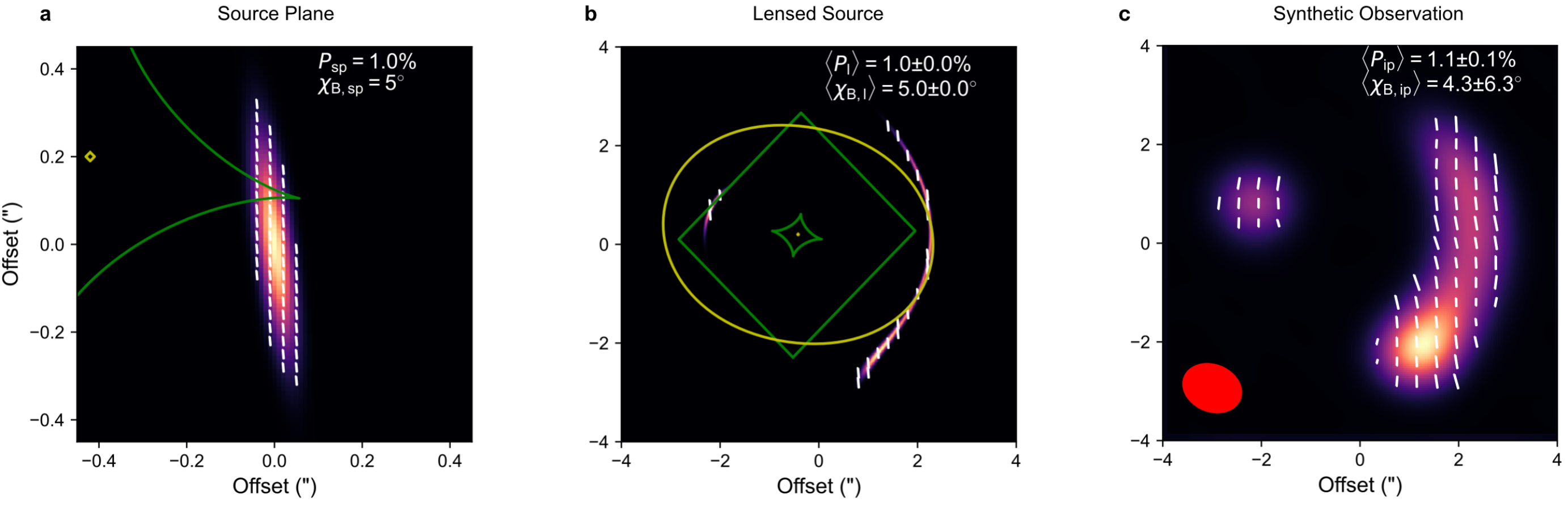}
    \caption{\textbf{Source plane configuration of the magnetic field and lensing model.} (\textbf{a}) Source plane intensity and field orientation. (\textbf{b}) Lensed source plane image. (\textbf{c}) Synthetic observations with the synthetic beam size ($1.2\arcsec\times0.9\arcsec$, $\theta=68$\,degrees) indicated by the red ellipse. The {\it B}-field orientation is indicated by white lines with lengths proportional to the polarization fraction. The median and root mean squared values of the polarization fraction and {\it B}-field orientation are indicated top right. The caustics in the source plane and image plane are shown as green and yellow lines respectively.}
    \label{fig:fig2}
\end{figure}

The mean and integrated polarization fractions of 9io9 are consistent with the $P\sim0.8$\% level measured in nearby spiral and starburst galaxies at wavelengths of 53--214\,$\mu$m\cite{Lopez-Rodriguez_2023}. The observations presented here are sensitive to polarized emission beyond this range, pushing into the Rayleigh-Jeans tail of the thermal emission spectrum at $\lambda_{\rm rest}= 350\,\mu$m. In recent models of diffuse interstellar dust, the polarization fraction, $P$, is independent of wavelength across 200--2000$\mu$m, consistent with observations of Galactic dust emission\cite{Hensley2022}. Observations of local starburst galaxies show that $P$ only varies by 0.4 per cent over the 50--150\,$\mu$m range, with an increase of up to $\sim$1 per cent towards 214\,$\mu$m\cite{Lopez-Rodriguez_2023}. We therefore conclude that 9io9 has a polarization level similar to local star-forming disks and starburst galaxies, with a key difference being the order-of-magnitude difference in gas mass and star-formation rate, with the disk of 9io9 being close to molecular gas dominated, contrasted with the $f_{\rm gas}\approx10$ per cent gas fractions of local star-forming disks\cite{Saintonge2022}. 

The large-scale ordered magnetic fields that exist in massive disk galaxies in the local Universe is thought to arise through the amplification of seed fields, and this has been predicted to occur on relatively short cosmological timescales, of order 1\,Gyr\cite{Beck1994, Brandenburg2005, Schober2013}. Weak seed fields (as low as $B\sim10^{-20}$\,G) could be formed in protogalaxies either through trapping of a cosmological field, possibly primordial in nature, or through the battery effect following the onset of star formation\cite{Biermann1950, Beck1996, Kulsrud2008, Subramanian2019}. Although turbulent gas motions in disks can reduce net polarization if they impart a strong turbulent component to the {\it B}-field\cite{Lee1985}, recent theoretical models of the formation of galactic-scale magnetic fields invoke turbulence in the ISM as the origin of a `small-scale' dynamo that can rapidly amplify the weak seed fields to $\mu$G levels\cite{Schober2013, Rieder2016, Rieder2017}. This small-scale dynamo is mainly driven by supernova explosions with coherence lengths of order 50--100\,pc, but turbulence can be injected into the ISM on multiple scales through disk instabilities and feedback effects, including stellar winds and outflows driven by radiation pressure, supernova explosions, and large scale outflows from an active galactic nucleus. 

The average turbulent velocity component of the disk of 9io9, determined from kinematic modelling of the CO emission, is $\sigma_v\approx70$\,km\,s$^{-1}$ and the star-formation rate density exceeds 100\,$M_\odot$\,yr$^{-1}$\,kpc$^2$\cite{Geach2018}. The high dense gas fraction of the molecular reservoir -- as traced by the ratio of CO(4--3)/C~{\sc i}(1--0) emission -- is also consistent with the injection of supersonic turbulence, which plays a key role in shaping the lognormal probability distribution function of the molecular gas density\cite{Padoan2002}. There is also tentative evidence of stellar feedback in action through the broad lines of dense gas tracers\cite{Geach2018}. Finally, one expects a high cosmic ray flux density in the ISM of 9io9, commensurate with the high star-formation rate density, and this too could serve to amplify magnetic fields. Therefore, 9io9 likely has the conditions required to rapidly amplify any weak seed fields via the small-scale dynamo effect, with amplification occurring on scales up to and including the full star-forming disk. Assuming equipartition between the turbulent kinetic and magnetic energies, we estimate an upper-limit of the equipartition turbulent {\it B}-field strength of 514~$\mu$G (Methods). This is comparable to the estimated turbulent {\it B}-field strength of $305\pm15~\mu$G within the central kiloparsec of the starburst region of M82 also using FIR polarimetric observations\cite{Lopez-Rodriguez_2021}. This indicates that the starburst activity of 9io9 could be be driving the amplification of {\it B}-fields across the disk.

Feedback-induced turbulence is a route to accelerating the growth of the seed fields, but to produce the ordered field on the kpc-scales observed requires a mean-field  dynamo\cite{Ruzmaikin1988,Beck1996}. This mean-field dynamo can be achieved through the rapid differential rotation of the gas disk, and this provides a mechanism for the ordering of an amplified {\it B}-field driven by star formation and stellar feedback processes. 9io9 is turbulent, intensely star-forming and rapidly rotating ($v_{\rm max}\approx300$\,km\,s$^{-1}$). This suggests that rather than an episode of violent feedback priming a large-scale but turbulent field that later evolves into an ordered field during a period of relative quiescence\cite{Rieder2017}, the small-scale and mean-field dynamo mechanisms operate in tandem. We estimate that the mean-field dynamo in 9io9 has not yet had time yto maintain or amplify the {\it B}-field (Methods). This indicates that the intense starburst is most important in amplifying the galactic field at $z=2.6$. We postulate that this `dual dynamo' might be the common mode by which galactic-scale ordered magnetic fields are established in young gas-rich, turbulent galaxies in the early Universe.  

Coherent magnetic fields consistent with the mean-field dynamo have been observed at $z=0.4$ via Faraday rotation of a background polarized radio source\cite{Mao17} (note that such observations are not possible for 9io9). Magnetic fields are already known to be present in the environment around normal galaxies at $z\approx1$ as revealed by the association of Mg~{\sc ii} absorption systems along quasar sightlines that exhibit Faraday rotation\cite{Bernet2008}, and indirectly through the existence of radio synchrotron emission from star-forming galaxies. However, mapping the \textit{B}-fields in individual galaxies at high redshift has so-far proven challenging. Our observations show that the polarized emission from magnetically aligned dust grains is a powerful tool to trace the {\it B}-fields of the cold and dense ISM in high redshift galaxies. 

9io9 is a particularly luminous example of a population of dusty star-forming galaxies in the early Universe that contribute a significant portion of the cosmic infrared background (CIB). If the one per cent level of polarization detected in 9io9 is representative of the general population of dusty star-forming galaxies\cite{Bonavera2017} then routine detection and mapping of magnetic fields in galaxies at high redshift is feasible (i.e.,\ in integration times of less than 24\,hr) even in unlensed systems with ALMA. This offers a new  window to characterise the physical conditions of the ISM in galaxies when galaxy growth was at its maximum, and will enable a better understanding the role of magnetic fields in shaping the early stages of galaxy evolution. The strength of the galactic magnetic field in local spiral galaxies is of order 10\,$\mu$G\cite{Beck2019}, and up to an order of magnitude higher in starbursts\cite{Lopez-Rodriguez2022}. Without resolving the polarization field in 9io9 below 100\,pc scales it is not possible to reliably estimate the {\it B}-field strength using dust polarization observations. Nevertheless, given the injection of kinetic turbulence driven by stellar feedback we estimate the strength of the {\it B}-field in 9io9 to be likely greater than that of local spiral galaxies, but similar to that of the central regions of nearby starburst galaxies (Methods).

Finally, these observations imply the CIB itself may be weakly polarized\cite{Bonavera2017,Feng2020}. Although misalignments of galaxies along the line of sight will serve to reduce the net polarization of the CIB, if the orientation of disks that host galactic-scale ordered {\it B}-fields is correlated on large scales due to tidal alignments\cite{Hirata2004}, then a polarization signal could remain and therefore fluctuations in the polarization intensity of the CIB could be used as a new probe of the physics of structure formation\cite{Feng2020}. This has consequences for cosmological experiments that seek to derive information on primordial conditions from observations of the polarization of the cosmic microwave background (CMB), especially if a curl component is present in the CIB polarization field\cite{Feng2020,Lagache2020}. A polarized component of the CIB at millimetre wavelengths, of extragalactic origin, dominated by emission at $z\approx2$ and with a power spectrum that is shaped by large scale structure at this epoch, will be a subtle but important foreground for future precision CMB experiments to contend with. 

\begin{addendum}
 \item This paper makes use of the following ALMA data: ADS/JAO.ALMA\#021.1.01461.S. ALMA is a partnership of ESO (representing its member states), NSF (USA) and NINS (Japan), together with NRC (Canada), MOST and ASIAA (Taiwan), and KASI (Republic of Korea), in cooperation with the Republic of Chile. The Joint ALMA Observatory is operated by ESO, AUI/NRAO and NAOJ. JEG and MD acknowledge support from the Royal Society.  ELR is funded by the Universities Space Research Association, Inc. (USRA) under the 07\_0032 NASA/DLR Stratospheric Observatory for Infrared Astronomy (SOFIA) program. GJB acknowledges support from STFC Grant ST/T001488/1.
 \item[Author Contributions]{JEG led the proposal to obtain the data and designed the observations; JEG and GJB reduced the data; JEG and ELR performed the analysis and all authors contributed to the manuscript.}

\end{addendum}


\clearpage

\section*{Methods}\label{app:methods} 

\subsection{Observations and data reduction}

9io9 at celestial coordinates $\alpha=02^{\rm hr}$09$^{\rm m}$41.3$^{\rm s}$, $\delta = +00^\circ$15$'$58.5$''$ (J2000) was observed by the ALMA 12\,m array in project 2021.1.01461.S. The target was observed in two sessions, where each session consisted of observations of sufficient duration to measure the rotation of the parallactic angle of the telescope by $\gtrsim$60\,degrees. The first session, comprising two Execution Blocks, was executed on 12 April 2022, and the second, comprising one Execution Block, was executed on 14 April 2022.  The ALMA grid source J2253+1608 was observed as a bandpass and flux calibrator, J0006$-$0623 was used as the polarization calibrator, J0208$-$0047 was used as the phase calibrator for the first session, and J0217+0144 was used as the phase calibrator for the second session.  

The data are processed in the Common Analysis Software Applications (CASA) version 6.2.1 using both the standard ALMA calibration pipeline and an additional polarization calibration script.  The standard calibration pipeline first applies a series of steps that include corrections to the amplitudes (based on system temperature measurements) and phases (based on water vapour radiometer measurements).  This is followed by calibration steps where corrections for the amplitudes and phases as a function of frequency are derived using the bandpass calibrator, scaling factors for the amplitudes are derived using the flux calibrator, and corrections for the phases and amplitudes versus time are derived using the phase calibrator.  The polarization calibration script derives and applies additional corrections for instrument-related polarization effects related to imperfections in the feeds and the orientation of the feeds as a function of time.  The uncertainty in the final linear polarization calibration fraction is 0.1 per cent, while the uncertainty in the polarization angle is 1\,degree.

We also use CASA for imaging of  the calibrated measurement set. The procedure {\it tclean} is used to produce the {\it IQUV} Stokes images, with the `{\it clarkstokes}' deconvolver, natural weighting of the visibilities and a common restoring beam (ensuring all Stokes images share the same synthetic beam). Dirty images are cleaned to a stopping threshold of 10\,$\mu$Jy and the final beam shape has a full width half maximum of $1.2''\times 0.9''$ at position angle 68\,degrees. All images are primary beam corrected to account for fall-off in sensitivity away from the phase centre, but note that the target is compact, and this represents a negligible correction to the measured flux densities. ALMA provides a systematic uncertainty in linear polarization of 0.03 per cent with minimum detectable polarization of 0.1 per cent, and therefore all quoted uncertainties in this work have the minimum detectable polarization of 0.1 per cent added in quadrature.

\subsection{Measuring polarization}

To account for the vector quantity of the polarization measurements, we estimate the integrated polarization fraction as

\begin{equation}\label{eq:P}
    P = \frac{\sqrt{\langle Q\rangle^{2} + \langle U\rangle^{2} - b}}{\langle I\rangle}
\end{equation}
\noindent
and the {\it B}-field orientation as
\begin{equation}\label{eq:PA_B}
   \chi_{B} = \frac{1}{2}\arctan{\left(\frac{\langle U\rangle}{\langle Q\rangle}\right)}+\frac{\pi}{2}
\end{equation}
\noindent
where $\langle I\rangle$, $\langle Q\rangle$, and $\langle U\rangle$ are the mean of the Stokes $I$, $Q$, and $U$ for measurements with a polarized intensity signal to noise ratio $P/\sigma_{P} \ge 3$, and bias $b$ is the standard error of the uncertainties of the Stokes {\it Q} and {\it U} images, $\sigma_{Q}$ and $\sigma_{U}$. The mean polarization fraction, $\langle P \rangle$, and {\it B}-field orientation, $\langle \chi_B \rangle$, are estimated as the mean distribution of the individual measurements of $p$ and $\chi_B$ in independent beams (i.e., using the individual Stokes $Q$ and $U$ and then computing $p$ and $\chi_B$). The uncertainties in these values are estimated as the standard deviation of each distribution, respectively. Extended Data Figure~1 shows the distribution of individual measurements at the Nyquist sampling for pixels with $PI/\sigma_{PI} \ge 3$.

\subsection{Lens modelling} 

Our goal is to obtain the  configuration of the {\it B}-field morphology in the source plane using previous knowledge of the lens model of 9io9. The lens model is computed using a singular isothermal ellipsoid (SIE) and a shear component. The lens is described by the critical radius (i.e.\ Einstein radius), $\theta_{E}$, lens offset position from the central coordinates, $x_{c}$, $y_{c}$, ellipticity, $\epsilon$, shear, $\gamma$, and shear orientation, $\theta_{\gamma}$. Extended Data Table~1 summarises the best-fit lens model parameters from$^5$. The morphology of the continuum thermal emission in the source plane is assumed to match that of the CO(4--3) molecular gas emission$^5$. The CO(4--3) spectral cube can be fit with a kinematic model based on the ring/disk morphologies of the molecular gas reservoirs of local ultraluminous galaxies\cite{Downes1998}, with the source plane spectral cube well-modelled with a disk of maximum radius $R_{\rm max}= 322^{+11}_{-20}$ mas ($2647^{+88}_{-160}$ pc) and maximum rotation speed of $V_{\rm max}=300$\,km\,s$^{-1}$). The disk has a tilt angle of $5\pm4$\,degrees (East of North) and is inclined along the line of sight by $50^{+3}_{-8}$\,degrees.

We produce synthetic polarimetric observations of our ALMA polarimetric observations as follows. 
First, we compute the SIE+shear lens model from the parameters given in Table~\ref{tab:tab1} using \textsc{lenstronomy}\cite{Birrer2018}. Using a grid of $801\times801$ pixels$^{2}$ with a scale of $0.01\arcsec$\,pixel$^{-1}$, which corresponds to a spatial scale of 82\,pc in the source plane, we model the source plane using an asymmetric 2D Gaussian profile with FWHM equal to the $R_{\rm{max}}$ and $R_{\rm{min}}$ at an angle of $\theta$ (Table \ref{tab:tab1}). The source plane and image plane at the native resolution of $0.01\arcsec$ is shown in Fig. \ref{fig:fig2}, also showing the caustics of the lens model. To mimic the observed data, we use the {\it simobserve} task of CASA, which simulates the observation of a given sky model (i.e.\ the modelled image plane) with the ALMA 12\,m array. We match the parameters of the real observations as closely as possible: we `observe' in two sessions using antenna configuration `2' for Cycle 8. The integrations start at hour angles of $-$2.339\,hours and $-$2.422\,hours for the two sessions and last 1.86 hours and 0.86 hours on 12 April 2022 and 14 April 2022, respectively, ensuring an identical sampling of the {\it uv} plane. A model is applied to simulate noise in the simulated observations, dominated by a thermal component, using the Atmospheric Transmission at Microwaves model to simulate the atmospheric profile at the ALMA site, with the precipitable water vapour (pwv) column as a scaling parameter. We assume a pwv of 0.6\,mm, equivalent to the average column across the observations. We clean the simulated visibilities in the same manner as the real data, but apply a restoring beam equivalent to the common beam derived for the real observations, to ensure that the final synthesised beams of the simulated images match the real data.

The {\it B}-field orientation in the source plane is assumed to be constant, $\chi_{B,{\rm sp}}$, with a constant polarization fraction of unity, $P_{\rm sp}$. The model Stokes $Q_{\rm sp}$ and $U_{\rm sp}$ in the source plane are computed as 

\begin{eqnarray}\label{eq:model_sp}
    Q_{\rm sp} = P_{\rm sp}\cos(2\chi_{B, {\rm sp}})\times I_{\rm sp} \\
    U_{\rm sp} = P_{\rm sp}\sin(2\chi_{B, {\rm sp}})\times I_{\rm sp} 
\end{eqnarray}
\noindent
where $I_{\rm sp}$ is the Stokes \textit{I} in the source plane. We multiply by $I_{\rm sp}$ to convert the Stokes $Q_{\rm sp}$ and $U_{\rm sp}$ in surface brightness, which allows us to compute the lens model. Figure~\ref{fig:fig2} shows the {\it B}-field orientation in the source plane with vector line lengths proportional to the polarization fraction, i.e.\ $P_{\rm sp}=1$\%. As the Stokes $Q_{\rm sp}$ and  $U_{\rm sp}$ are density profiles, these images can also be lensed using the same procedure as the Stokes $I_{\rm sp}$. 

The polarized intensity, $PI_{\rm ip}$, polarization fraction, $P_{\rm ip}$, and {\it B}-field orientation, $\chi_{B, {\rm ip}}$, of the final synthetic polarimetric observation in the image plane are computed as

\begin{eqnarray}\label{eq:mock_pol}
    PI_{\rm ip} = \sqrt{Q_{\rm ip}^2 + U_{\rm  ip}^{2}} \\
    P_{\rm ip} = \frac{PI_{\rm ip}}{I_{\rm ip}} \\
    \chi_{B, {\rm ip}} = \frac{1}{2}\arctan{\left(\frac{ U_{\rm ip}}{Q_{\rm ip}}\right)} + \frac{
\pi}{2}
\end{eqnarray}
 
\noindent The final synthetic observations Stokes \textit{I$_{\rm ip}$}, \textit{Q$_{\rm ip}$}, and \textit{U$_{\rm ip}$} are shown in Figure~\ref{fig:fig1}. This figure also shows the synthetic polarized intensity, $PI_{\rm ip}$, and the {\it B}-field orientation, $\chi_{B,{\rm ip}}$, with the length of the lines proportional to the synthetic polarization fraction $P_{\rm ip}$. Note that the {\it B}-field orientation of the synthetic observations does not have the added rotation of $90$\,degrees as shown in Equation~\ref{eq:PA_B}. This is because we are modeling the {\it B}-field rather than the {\it E}-vector measured by the ALMA polarimetric observations. However, to compare with our observations, Figure~\ref{fig:fig1} shows the synthetic Stokes \textit{Q} and \textit{U} in the same reference (i.e.\ {\it E}-vector) as the ALMA polarimetric observations.

\subsection{Constraining the {\it B}-field orientation}\label{app:BPA}

It has been shown that for a non-rotating lens the polarization vector of the electromagnetic wave does not rotate. This result has been demonstrated using pure geometrical definitions\cite{Dyer1992,Burns2002}, and from first principles using a Newtonian potential and solving the spacetime metrics\cite{Faraoni1993}. In general, the photons travel along null geodesics, which implies that the vector properties are time invariant across the geodesic. This property allow us to study the intrinsic {\it B}-field geometry of the source amplified by the gravitational lens.

We explore potential changes of the {\it B}-field orientation in our lens model. With a polarization fraction of 1\% we set a range of constant $\chi_{B,{\rm sp}}$ orientations in the source plane. Extended Data Figure~2 shows the source plane and lensed model for several values of $\chi_{B, {\rm sp}} = 5, 10, 45, 90$\,degrees at a constant $P_{\rm sp} =1$\% using the same lens model as described above, and simulating the observations in the same way to produce convolved and noisy images at the same scale as the data. We conclude that the lens model and convolution do not produce rotation or change in the polarization fraction from the source plane to the image plane. For $\chi_{B,{\rm sp}} = [5,10]$\,degrees, the final $B_{\rm{ip}}$ has an orientation similar to the beam PA in regions with low SNR. However, the median {\it B}-field orientation in the final synthetic observation is consistent with the modeled \textit{B} field orientation in the source plane, with the presence of noise and asymmetric synthetic beam contributing to the angular dispersion in the image plane. We find that the background noise and {\it uv} plane sampling (in general, imaging of the visibilities) introduce an uncertainty of $\sim10$\,degrees in $\langle \chi_{B,{\rm ip}} \rangle$, where the uncertainty is dominated by the signal-to-noise of the ALMA polarimetric observations. 

The \textit{B}-field orientation can also be constrained using the Stokes $Q$ and $U$ images. Our observations show that Stokes $Q$ is negative and Stokes $U$ is consistent with zero (Figure~\ref{fig:fig1}). This result shows that the {\it E}-vector is mainly in the north-south direction. This configuration gives us the opportunity to tightly constrain the possible range of values in the {\it B}-field orientation. Extended Data Figure~3 shows the Stokes $I$, $Q$, and $U$ images, and the polarized intensity of synthetic observations for the same set of artificial \textit{B}-field orientations in the source plane. Note the change from negative to positive $Q$ from $\chi_{B, {\rm sp}}$ of $5$\,degrees to $90$\,degrees. 
This behaviour is expected, but we show it for completeness. We can constrain the orientation by rotating the \textit{B}-field until the corresponding Stokes $U$ image shows a $3\sigma$ detection in the synthetic polarimetric observations. This condition is met when $\chi_{B, {\rm sp}} = \pm10$\,degrees. For orientations deviating outside $\pm$$10$\,degrees the synthetic observations are not consistent with our observations. We conclude that the most likely {\it B}-field orientation consistent with our observations is $\chi_{B,{\rm sp}} = 5_{-15}^{+5}$\,degrees, where we have set the {\it B}-field orientation to be parallel to the CO(4--3) emission of the disk in the source plane$^5$. 

\subsection{Constraining the polarization fraction}

We assume a constant polarization fraction of 1 per cent in the source plane. The lens model shows that the median polarization fraction is consistent with this level without  dispersion (Extended Data Figure~2). We conclude that the lens model does not change the polarization fraction from the source plane to the image plane. Using the simulated observations, we estimate the median and r.m.s.\ polarization fraction, finding that the combination of {\it uv}-plane sampling and background noise adds an uncertainty of $\sim$0.1--0.2 per cent in the polarization fraction. 

\subsection{Energy equipartition}

We estimate the {\it B}-field strength assuming equipartition between the turbulent kinetic energy and the turbulent magnetic energy. Let the turbulent kinetic energy, $U_{K}$, and turbulent magnetic energy, $U_{B}$, be

\begin{eqnarray}
    U_{K} &=& \frac{1}{2} \rho \sigma_{\rm{v}}^{2} \\
    U_{B} &=& \frac{B^{2}}{8\pi}
\end{eqnarray}

\noindent where $\rho$ is the volume density, $\sigma_{\rm{v}}$ is the velocity dispersion, and $B$ is the field strength. Then, assuming equipartition, $U_{K} = U_{B}$, the field strength is 

\begin{equation}
    B_{\rm{eq}} = \sqrt{4\pi\rho}\sigma_{\rm{v}}.
\end{equation}

To estimate the baryonic volume density, $\rho = M/V$, we use the disk volume, $V = \pi r^2 h$, and mass of the molecular gas of the galaxy. The radius is $r =2647^{+88}_{-160}$ pc and the height is $h \le 600$ pc$^5$; note that the disk height is unresolved in the observations of the CO molecular gas, so we take the size of the beam of the observations as an upper limit, also noting that the scale heights of bursty disk galaxies at high redshift are observed to be larger than local disks, with heights of several hundred parsecs typical. The total molecular mass is estimated from the CO luminosity$^5$ as $M_{\rm{H_{2}}} = (7.5\pm0.1)\times10^{10}$ M$_{\odot}$ and the velocity dispersion of the gas is $\sigma_{\rm{v}} = 73\pm4$ km s$^{-1}$, from kinematic fitting of the disk model to the CO data cube$^5$. From this, we estimate an equipartition {\it B}-field strength of $B_{\rm{eq}} \le514~\mu$G, however note that the observed velocity dispersion may be affected by large-scale flows from galactic winds and shearing effects by the rotating galactic disk, such that the turbulent component of the velocity dispersion might be lower. These effects produce an overestimated measurement of the {\it B} field strength\cite{Guerra2023}.

Spiral galaxies have an average ordered {\it B}-field strength of around $5\pm2$\,$\mu$G with a total {\it B}-field strength of $17\pm14$~$\mu$G assuming equipartition between the total {\it B}-field and total cosmic-ray electron density$^1$. A revised equipartition formula to account for energy losses in nearby (within 160\,Mpc) starburst galaxies estimated\cite{LB2013} equipartition {\it B}-field strengths in the range $70-770~\mu$G. Recent FIR polarimetric observations$^{21}$ of the starburst galaxy M82 showed that the turbulent {\it B}-field strength is $305\pm15~\mu$G within the central kiloparsec region. These authors modified the DCF method to account for the large-scale flow of the galactic outflow. The correction from the equipartition {\it B}-field strength, $B_{\rm eq}=540\pm170~\mu$G, was estimated to be $\sim$25\%. 

\subsection{Ordering timescale}

We estimate the timescale to order a large-scale magnetic field in a disk. The ordering timescale can be estimated as\cite{Arshakian2009}

\begin{equation}
    \tilde{t} = \frac{h^{2} l_{\rm{c}}}{\beta h} |D_{\rm{d}}|^{-1/4}
\end{equation}
\noindent
\noindent where $h$ is the half-thickness of the galactic disk and $l_{\rm{c}}$ is the large-scale coherence length. $\beta$ is the turbulent diffusivity defined as $\beta=l\sigma_{\rm v}/3$, where $l$ is the coherence length of the small-scale turbulence, and $\sigma_{\rm v}$ is the turbulent velocity. $D_{\rm{d}}$ is the dynamo number defined as

\begin{equation}
|D_{\rm{d}}| = 9 \left( \frac{h\Omega}{\sigma_{\rm v}} \right)^{2}
\end{equation}
\noindent
\noindent where $\Omega$ is the angular velocity.

For 9io9 we assume a Gaussian scale height$^5$ $h=300$\,pc and $\sigma_{\rm v}=73\pm4$ km s$^{-1}$ of the molecular gas. To estimate the angular velocity we use the de-projected circular velocity, $v_{\rm{max}} = 360^{+49}_{-11}$ km s$^{-1}$, at the maximum radius of the disk, $r_{\rm max} = 2647^{+88}_{-160}$ pc, yielding $\Omega = v_{\rm{max}}/r_{\rm{max}} = 145\pm8$ km s$^{-1}$ kpc$^{-1}$. 

The typical coherence length of the small-scale dynamo driven by stellar activity in the molecular gas of galaxies on scales of $l=1$--$10$\,pc\cite{Haverkorn2008}. We assume a large-scale coherence length in the range $l_{\rm{c}} = 0.5-2.0$\,kpc, corresponding to $50-2000$ times larger than the turbulence coherence length, $l$. We estimate a dynamo number $D_{\rm{d}} = 3.2\pm0.4$ and turbulent diffusivity $\beta= (4.2\pm2.0) \times 10^{25}$ cm$^{2}$ s$^{-1}$.  Finally, the ordering timescale is estimated to be in the range of $\tilde{t} \approx 0.4$--$18$\,Gyr, noting that only 2.5\,Gyr have elapsed by $z=2.6$. Analytical solutions of the evolution of \textit{B}-fields  in spiral galaxies predicts that a large-scale \textit{B}-field can be formed within $2$--$5$ Gyr \cite{Rodrigues2019}. This study suggests that the mean-field dynamo may be already active at $z<3$. Note that 9io9 has very high star formation rate, which may inject higher turbulent energy into the system than those studied in ref 37 affecting the ordering of the {\it B}-field in the galaxy's disk. The regular {\it B}-field is amplified if the dynamo number is larger than the critical value of $D_{\rm{d,cr}} \sim 7$ estimated from numerical simulations of galactic dynamo models$^{22}$. For comparison, the Milky Way has a dynamo number of $D_{\rm d} \sim 9 > D_{\rm{d,cr}}$, which shows that large-scale dynamo mechanism is important in our Galaxy. For 9io9 we estimate $D_{\rm d} = 3.2\pm0.4$, below the critical value. In addition, the time for 9io9 to complete a rotation is $t\sim45$\,Myr, which indicates that approximately $9$--$400$ galactic rotations at a galactocentric radius of $2.6$\,kpc are required to order the large-scale {\it B}-field.


\newcounter{mybibstartvalue}
\setcounter{mybibstartvalue}{27}

\xpatchcmd{\thebibliography}{%
  \usecounter{enumiv}%
}{%
  \usecounter{enumiv}%
  \setcounter{enumiv}{\value{mybibstartvalue}}%
}{}{}

\begin{addendum}
 \item[Data Availability] The ALMA data used in this work will be available via the ALMA Science Archive (https://almascience.nrao.edu/aq) with reference to the project code 2021.1.01461.S.
 \item[Competing Interests] The authors declare that they have no competing financial interests.
 \item[Author Information] Correspondence and requests for materials
should be addressed to J.\,E.\,Geach~(email: j.geach@herts.ac.uk).
\end{addendum}
\clearpage

\section*{Extended Data}\label{app:extdata} 

\setcounter{figure}{0}    

\renewcommand{\tablename}{ED Table}

\begin{table}
\begin{center}
\begin{tabular}{lcc}
Einstein radius 		& $\theta_{\rm E}$		& 2.45$\arcsec$	\\		
Ellipticity 			& $\epsilon$ 			& 0.072  \\
Ellipticity angle 		& $\theta_{\epsilon}$ 	& 67.2$^{\circ}$  \\
Shear 			        & $\gamma$ 			    & 0.056  \\
Shear orientation 	    & $\theta_{\gamma}$ 	& $-$91$^{\circ}$  \\
Right Ascension 	& $x_{c}$				& 02$^{\text{h}}$09$^{\text{m}}$41.254$^{\text{s}}$ \\
Declination	& $y_{c}$				& $+$00$^{\circ}$15$^{'}$58.450$^{''}$ \\
\end{tabular}
\caption{Lens model parameters}\label{tab:tab1}

\item Angles are measured counterclockwise, i.e., East of North.
\item Coordinates in J2000.
\end{center}
\end{table}

\renewcommand{\figurename}{ED Figure}
\begin{figure}[h]
    \centering
    \includegraphics[width=\textwidth]{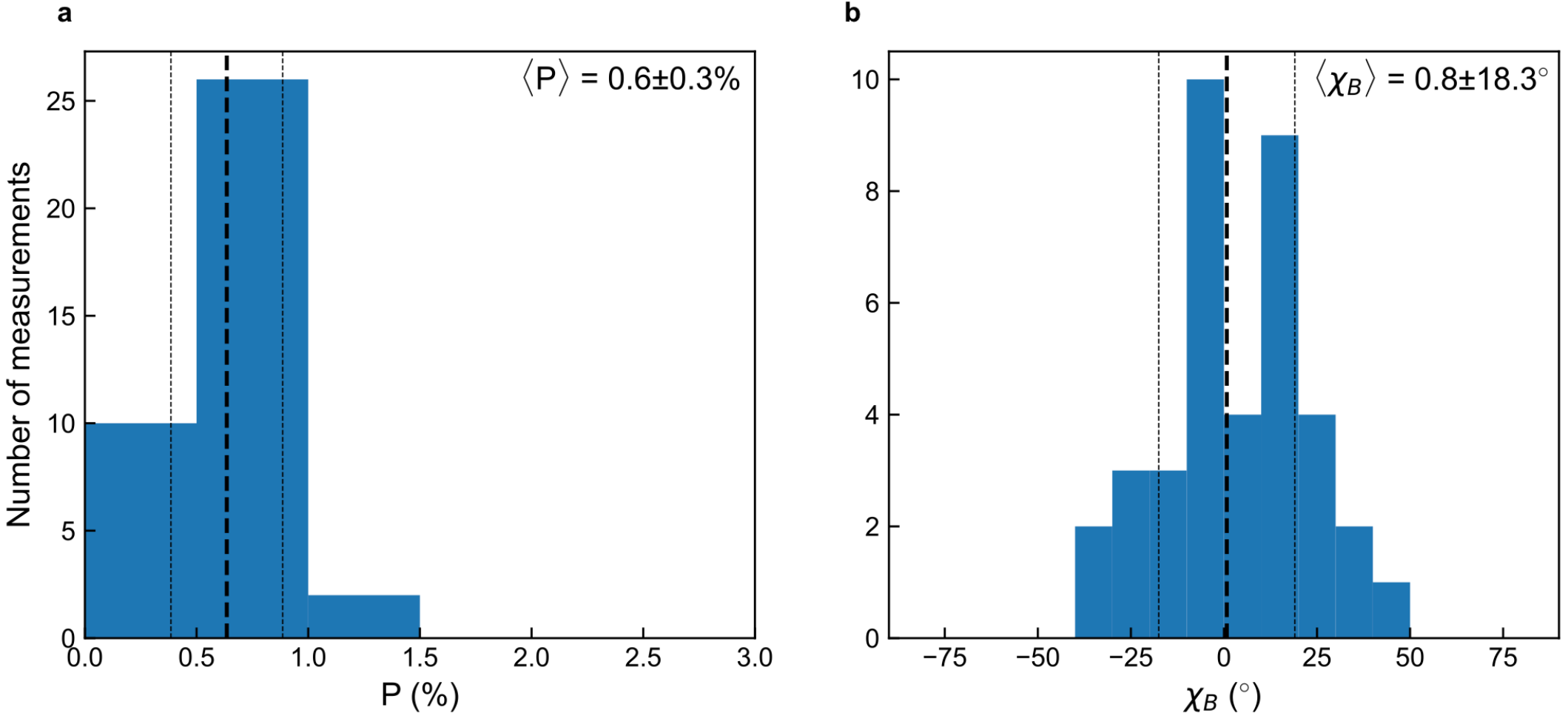}
    \caption{{\bf Histograms of the measured polarization fraction and magnetic field orientation.} The polarization fraction (\textbf{a}) is displayed in bins of $0.5$\% and the {\it B}-field orientation (\textbf{b}) in bins of 10\,degrees, where individual measurements are taken in independent beams. The median and $1\sigma$ scatter of the distributions is given for each measurement.}
    \label{fig:pol_mes}
\end{figure}

\begin{figure}
    \centering
    \includegraphics[width=\textwidth]{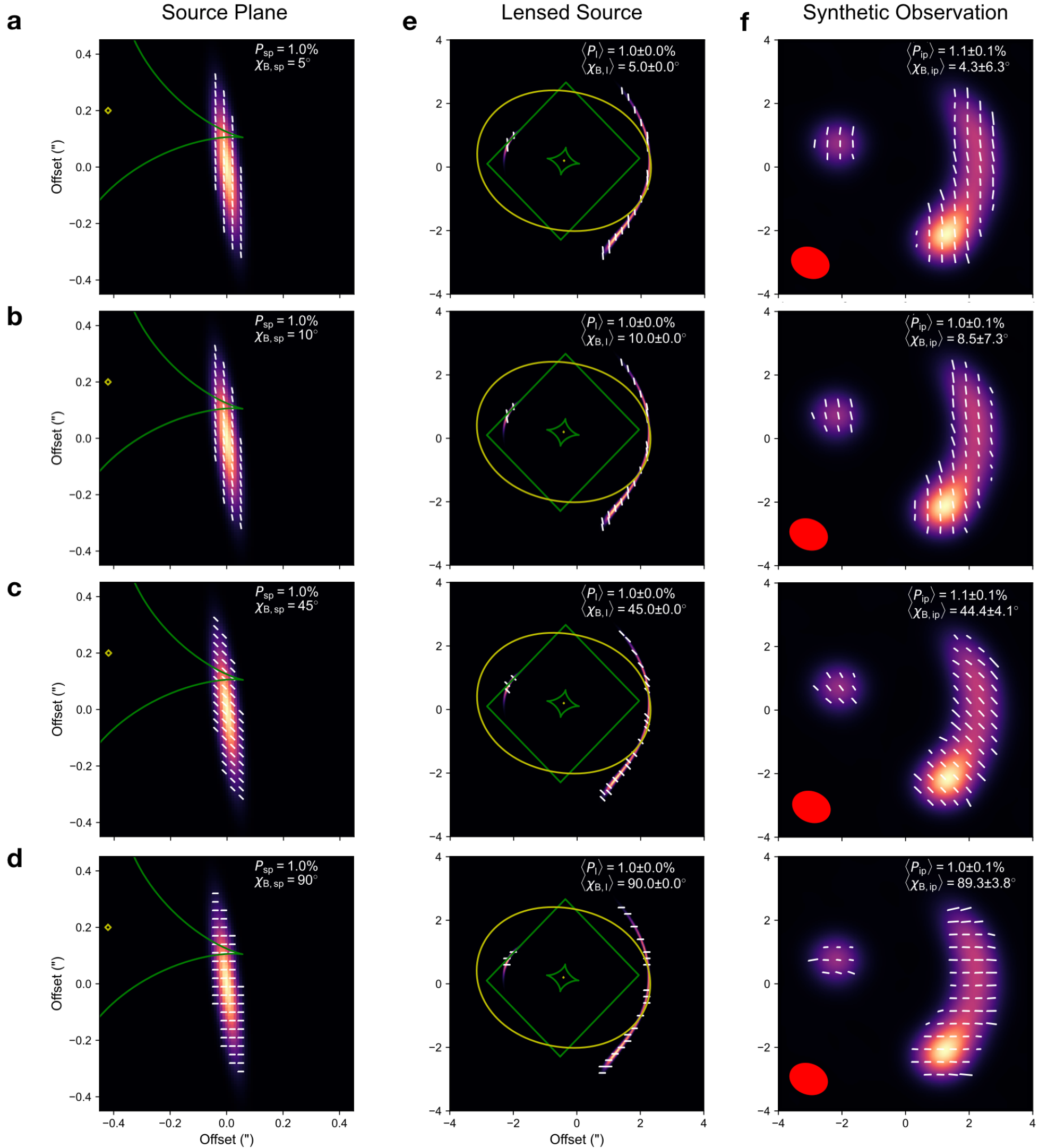}
    \caption{\textbf{Characterisation of the magnetic field orientation and polarization fraction.} The \textit{B}-field orientation (white lines) at 5\,degrees (\textbf{a}), 10\,degrees (\textbf{b}), 45\,degrees (\textbf{c}), and 90\,degrees (\textbf{d}) are shown over the source plane (\textbf{a}), lens model (\textbf{e}), and synthetic observations (\textbf{f}). We indicate the caustics in the source plane (green line), and image plane (yellow line). The median and r.m.s.\ of the polarization fraction and \textit{B}-field orientation are shown at the top right of each panel. The red ellipse indicates the synthetic beam of the simulated observations.}
    \label{fig:fig4}
\end{figure}

\begin{figure}
    \centering
    \includegraphics[width=\textwidth]{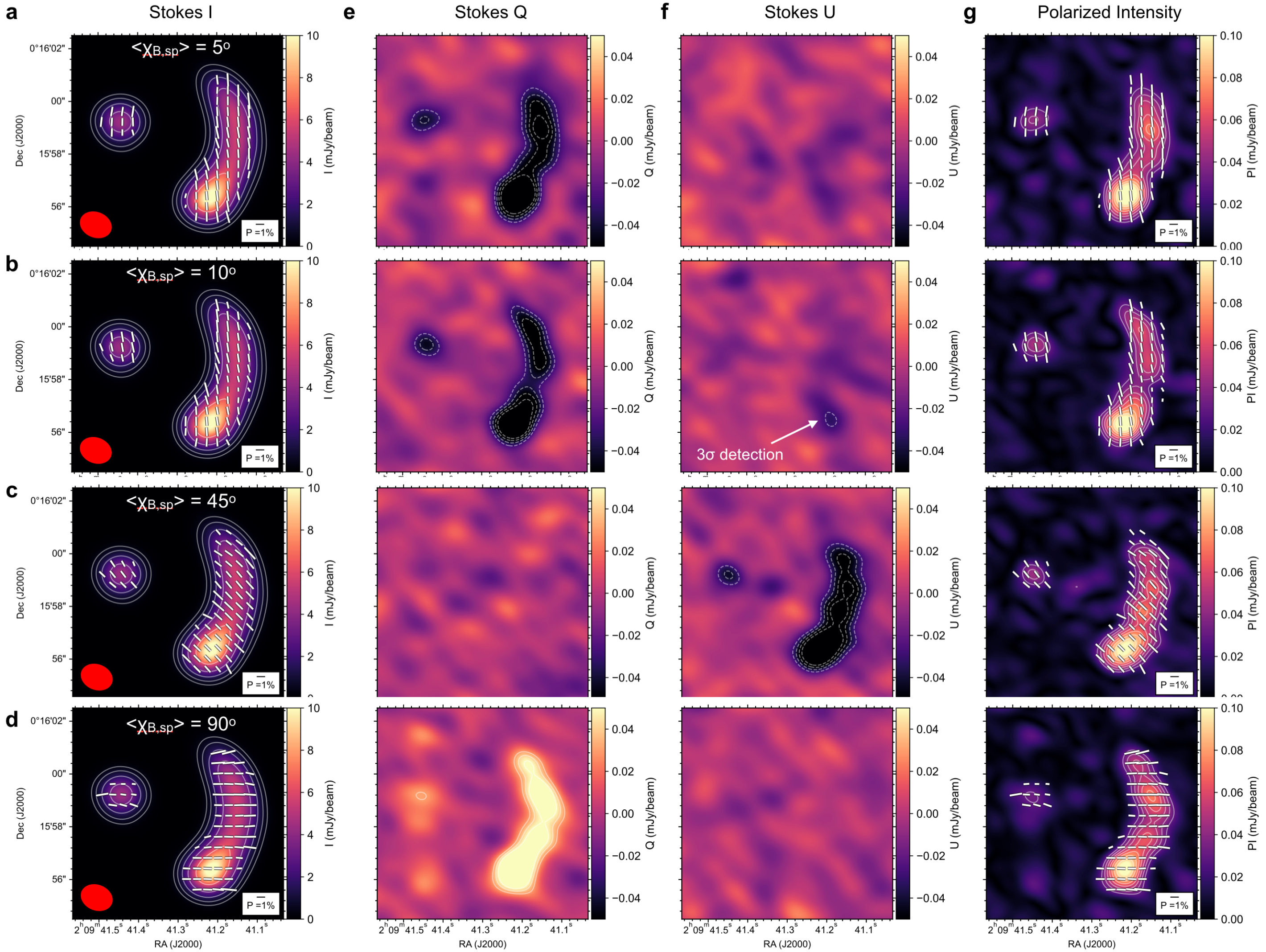}
    \caption{\textbf{Variation of the Stokes \textit{Q} and \textit{U} parameter with magnetic field orientation}. The \textit{B}-field orientation (white lines) at 5\,degrees (\textbf{a}), 10\,degrees (\textbf{b}), 45\,degrees (\textbf{c}), 90\,degrees (\textbf{d}) are shown over the Stokes $I$ (\textbf{a}), Stokes $Q$ (\textbf{e}),  Stokes $U$ (\textbf{f}), and polarized intensity (\textbf{g}) of the  synthetic polarimetric observations.  The FWHM of the observations is shown at the bottom-left (\textbf{a}). For Stokes $I$, the contours increase as $\sigma_{I} \times 2^{3,4,5, \dots}$. For Stokes $Q$ and $U$, and $PI$, the contours start at $3\sigma$ and increase in steps of $1\sigma$. Note that at $\sigma_{B,{\rm sp}} = 10$\,degrees a $3\sigma$ signal becomes detectable in Stokes $U$.}
    \label{fig:fig5}
\end{figure}

\end{document}